# Gate Tunable Magnetism and Giant Magnetoresistance in ABC-stacked Few-Layer Graphene

Yongjin Lee[1,2], Shi Che[3], Jairo Velasco Jr.[4], David Tran[2], Jacopo Baima[5], Francesco Mauri[6], Matteo Calandra[5*], Marc Bockrath[3], Chun Ning Lau[3*]

[1] Samsung Semiconductor R&D center, Hwasung-si, Gyeonggi-do, South Korea.
[2] Department of Physics and Astronomy, University of California, Riverside, Riverside, CA 92521
[3] Department of Physics, The Ohio State University, Columbus, OH 43210
[4] Department of Physics, University of California, Santa Cruz, CA 95064
[5] Sorbonne Université, CNRS, Institut des Nanosciences de Paris, UMR7588, F-75252, Paris, France
[6] Dipartimento di Fisica, Università di Roma La Sapienza, Piazzale Aldo Moro 5, I-00185 Roma, Italy

**Magnetism is a prototypical phenomenon of quantum collective state, and has found ubiquitous applications in semiconductor technologies such as dynamic random access memory (DRAM). In conventional materials, it typically arises from the strong exchange interaction among the magnetic moments of *d*- or *f*-shell electrons. Magnetism, however, can also emerge in perfect lattices from non-magnetic elements. For instance, flat band systems with high density of states (DOS) may develop spontaneous magnetic ordering, as exemplified by the Stoner criterion. Here we report tunable magnetism in rhombohedral-stacked few-layer graphene (r-FLG). At small but finite doping ($n\sim10^{11}$ cm$^{-2}$), we observe prominent conductance hysteresis and giant magnetoconductance that exceeds 1000% as a function of magnetic fields. Both phenomena are tunable by density and temperature, and disappears for $n>10^{12}$ cm$^{-2}$ or $T>5$K. These results are confirmed by first principles calculations, which indicate the formation of a half-metallic state in doped r-FLG, in which the magnetization is tunable by electric field. Our combined experimental and theoretical work demonstrate that magnetism and spin polarization, arising from the strong electronic interactions in flat bands, emerge in a system composed entirely of carbon atoms. The electric field tunability of magnetism provides promise for spintronics and low energy device engineering.**

The advent of van der Waals materials has provided platforms for investigating magnetism with unprecedented controllability, as shown by recent works on magnetic monolayers such as CrI$_3$ and Fe$_3$GeTe$_2$[1,2,3,4]. An exciting development in this frontier is the emergence of magnetism in flat band systems with non-magnetic elements, such as that in twisted bilayer graphene[5], and the prospect of electrical control of magnetic orders.

Rhombohedral-stacked few-layer graphene (r-FLG) is a 2D material with exceedingly flat bands. In the absence of interactions, calculations predict that r-FLG should host a surface band dispersion that scales approximately with $k^N$, where $N$ is the number of layers, leading to flat bands and diverging density of states (DOS) near the K points[6,7]. Due to the instability of the large DOS to electronic interactions, the ground state of charge neutral r-FLG may not be an isotropic Fermi liquid, but phases with spontaneously broken symmetries[8-24]. Various magnetic instabilities, such as ferromagnetism and antiferromagnetism have been proposed[21,25-29], though apart from a possible antiferromagnetic state at the charge neutrality[30-33], no evidence of magnetism has been reported.

---

* Email: calandra@insp.jussieu.fr, lau.232@osu.edu,

Here we focus on suspended r-FLG devices, which, due to the absence of substrate-induced screening, host even stronger electronic interactions than devices supported on hexagonal BN or $SiO_2$ substrates. A typical device image is shown in Fig. 1a inset. **Fig. 1a** displays a rhombohedral-stacked trilayer graphene (r-TLG) device's two terminal differential conductance $G=dI/dV$ as a function of source-drain bias $V$ at magnetic field $B=0$ at the charge neutrality point (CNP). $G=0$ for small $V<40$ mV, but displays extremely sharp peaks at $V=\pm42$mV, indicating the presence of an interaction-induced energy gap that is ~40 meV in magnitude. This insulating state persists for gate voltages $|V_g|<1.65$V (**Fig. 1b**), and can be suppressed by increasing temperature $T$, by an out-of-plane displacement field $D$, or by a parallel magnetic field $B_\parallel$[30,31]. This gapped insulating phase, which appears to be the ground state of charge neutral r-TLG, is most consistent with a layer antiferromagnetic state with spontaneously broken time-reversal symmetry, where the electrons on the top and bottom layers are oppositely spin polarized (Fig. 1c), and the middle layer has substantially smaller magnetic moments.

We now explore magnetotransport of r-TLG at finite doping in both parallel and perpendicular magnetic fields ($B_\parallel$ and $B_\perp$). For $|V_g|>1.65$V or charge density $n>\sim7.5\times10^{10}$ cm$^{-2}$, r-FLG is conductive. Strikingly, its conductance displays prominent hysteresis as a function of an in-plane magnetic field $B_\parallel$. Fig. 1d displays $G(B_\parallel)$ at $V_g=12.2$ V and $T=0.3$K: as $B_\parallel$ sweeps up (down) from -4T (4T), $G$ decreases steadily; it rises sharply when $B_\parallel$ crosses 0, reaching a maximum at ~ 1.5T (-1.5T) before decreasing again. The up- and down-sweep curves $G(B_\parallel)$ are approximately mirror symmetric with respect to $B_\parallel$, and are reproducible upon repeated sweeps. Similar hysteresis loops are observed in 5 different r-TLG and 4LG devices, but are minimal in Bernal-stacked bilayer or trilayer graphene samples that are similarly fabricated and measured. Thus the hysteresis does not arise from experimental or instrumental artifacts, but instead appears to be intrinsic to r-FLG samples at intermediate densities.

The hysteresis loops are strongly reminiscent of the giant magnetoresistance effect observed in magnetic multilayers[34] – in fact, the ground state of r-TLG bears similarity with a magnetic multilayer, as it consists of two surface layers with intra-plane antiferromagnetic couplings, separated by middle layers formed by atoms with substantially smaller magnetic moments. To further explore the hysteresis, we examine the evolution of normalized magnetoconductance (MC) with $n$ and $T$, where MC is defined as $(G(B)-G(0))/G(0)$. **Fig. 2a** displays MC($B_\parallel$) at $T=0.3$K for up and down sweeps at different densities, $V_g=-4.8$, 12.2 and -11.3 V, respectively. Very large MC is observed, reaching as much as 1000 at low charge density, $|V_g|=\sim3.8$V; accompanying this giant MC is prominent hysteresis with a characteristic butterfly shape. The upsweep and downsweep curves coincide at $|B_\parallel|>3.5$T, which we take as the saturation field $B_{sat,\parallel}$. As the device become more electron- or hole-doped, MC decreases quickly. At $V_g<-15$V, MC is reduced to a few percent, while hysteresis almost completely vanishes. At all temperatures, MC is negative at small $B_\parallel$, and becomes positive at large $B_\parallel$.

Similarly prominent MC and hysteresis are also observed as a function of out-of-plane magnetic field $B_\perp$ (Fig. 2c). Similar to the case of parallel field, they are suppressed by higher temperature and doping, though the saturation field $B_{\perp,sat}$ is ~2T. Another difference is that in perpendicular field, MC is positive at low temperatures, but becomes negative at $T=4.2$K, which we attribute to weak localization to weak antilocalization transition.

Notably, the magnitude of MC and hysteresis are also strongly temperature dependent. As shown by Fig. 2b and 2d, which plot MC at $V_g$~-4V at temperatures ranging from 0.28K to 4.2K, the giant MC and strong hysteresis loop at $T$=0.28K are suppressed by increasing temperatures. At 4K, MC~1% at 4K, and hysteresis is barely discernible. The quantitative temperature dependence of MC on temperature are exhibited in Fig. 3a-b, which plot MC($T$) at $B_\parallel$=1.5 T and $V_g$=-3.8V in linear and Arrhenius plots, respectively. The data can be fitted to thermal activation over an energy barrier of characteristic temperature ~ 4.5K.

To analyze the hysteresis loop's dependence on temperature, we quantify its amplitude by $\Delta G(B)=G_{up}(B)-G_{down}(B)$, where $G_{up}$ and $G_{down}$ refers to data obtained from up- and down-sweeps, respectively. Typical $\Delta G(B)$ at $V_g$=-3.8V and $T$=0.3, 1.8 and 5K are shown in Fig. 3c inset. For constant $B_\parallel$, $\Delta G(B)$ decreases approximately exponentially in temperature, with a characteristic energy scale ~ 1K (Fig. 3c). $B_{max}$, the magnetic field at which the hysteresis is maximum, appears to decrease linearly with temperature. In GMR structures, $B_{max}$ is close to the coercive field, which diminishes by raising temperature[35-37], qualitatively resembling our results. Finally, the presence of substantial hysteresis both in parallel and perpendicular magnetic field suggests that trilayer ABc graphene is a soft magnet with a low coercivity, i.e. the weakness of spin-orbit interaction in Carbon does not lead to an easy axis. As a result the weak magnetic moments can be easily oriented by the external magnetic fields. This is confirmed by first principles calculations showing magnetic moments smaller than 0.5% of a Bohr magneton.

The extremely large magnetoresistance, together with the hysteresis, suggests the presence of intrinsic magnetic ordering in r-TLG. In order to investigate the possible occurrence of magnetism and its dependence on doping, we perform first principles calculations for trilayer ABC graphene in field effect configuration[38,39]. At zero doping, r-TLG is found to be a perfect insulating antiferromagnet with degenerate up and down "mexican hat"-shaped bands. As shown in Fig. 4 (a), the PBE0 functional gives gaps of the order of 40 meV, in good agreement with experimental data. The magnetic moments are larger in the outermost layers ($\mu$=0.005$\mu_B$) and smaller in the internal one ($\mu$=0.003$\mu_B$), mainly because the top of valence and conduction bands are essentially formed by $2p_z$ states of two of the four carbon atoms in the outermost layer[40]. This is the so-called LAF state that preserves inversion symmetry[30]. The introduction of doping in FET configuration via an electric field perpendicular to the layer breaks the inversion symmetry, removes the spin-degeneracy and shifts the Fermi level in the conduction band. As a result, the bands become spin-split: the conduction band with ↑-spin states from the bottom layer has lower energy than that of the ↓ spin states from the top layer; in the valence band, the ↓-spin states from the bottom layer has higher energy (Fig. 4a). As the Fermi level is raised (lowered) to reside between the spin-split conduction (valence) bands, r-TLG is expected to be half-metallic with 100% spin polarization. Upon heavy electron- or hole-doping, the Fermi level crosses the other spin channel and half-metallicity is lost, reverting the device into a paramagnet. The evolution of the magnetic state and of the half-metallic state as a function of doping is shown in the phase diagram of Fig. 4d. These predictions agree well with our experimental observation (Fig. 4e): hysteresis and large magnetoresistance is only observed at finite intermediate doping, between $1\times10^{11}$ and $7\times10^{11}$ cm$^{-2}$, and disappears at higher doping. In theory, the domain of stability of the half-metallic and magnetic solutions extend up to $8\times10^{11}$ cm$^{-2}$, in quantitative agreement of what found in experiments. Thus, both theory and experiments confirm the occurrence of a magnetic ground state in trilayer ABC graphene.

In conclusion, by performing transport experiments and first principles electronic structure calculations we have shown that trilayer ABC graphene displays gate tunable intrinsic magnetism, prominent hysteresis in magnetoconductance and giant magnetoresistance, which is indicative of a layer antiferromagnetic magnetic ground state and a soft magnetic behavior with low coercivity. Our work provides evidence of *sp* magnetism in the flat band of few-layer graphene with rhombohedral stacking. The presence of a magnetic hysteresis at finite doping is suggestive of the half-metallic state predicted in ref. 29, and its unequivocal confirmation necessitates further experimental and theoretical efforts.

## Materials and Methods
### Sample Preparation
Rhombohedral-stacked trilayer and tetralayer graphene sheets are exfoliated from micromechanical cleaving of graphite crystals onto $SiO_2$/Si substrates, and identified by optical contrast and Raman spectroscopy. Multiple steps of electron beam lithography are employed to fabricate the electrodes and, in some devices, air-bridge top gates[41]. Buffered oxide etch is used to partially remove the $SiO_2$ layer underneath r-TLG sheets, and the completed devices are dried in a critical point dryer to minimize the surface tension of liquids. Typical gate capacitance $C_g$ of a device is approximately 50-80 $\mu F/m^2$. All data are taken using standard lock-in techniques in a $He^3$ refrigerator unless otherwise specified.

### First principles calculations.
We use a Def2-TZVP basis set with an added diffuse p function[42], reoptimized on monolayer graphene. In order to avoid numerical instabilities due to an ill-conditioned overlap matrix (basis set near-linear dependency), we removed the eigenvectors belonging to the smallest eigenvalues with a thereshold of $10^{-5}$. We use the hybrid functional PBE0 within density functional theory (DFT) and an all-electron localized Gaussian basis set (CRYSTAL code[43]) as this method was shown to provide reliable results for the stabilization of the magnetic state[40]. Sampling of reciprocal space is performed on mesh of 512x512 points, while the tolerance for the computation of direct space electrostatic and exchange summations are set to 8 8 8 15 30. The field effect is simulated by using a uniform layer of point charges at a distance of 3.35 Å from the trilayer below each hollow site of the graphene layer. In order to avoid spurious dipole-dipole interactions and the need of introducing a compensating dipole potential[29, 44], we do not impose periodic boundary condition along the z-axis[43].

## Acknowledgement
The experiments are supported by DOE BES Division under grant no. DE- SC0020187. Device fabrication is partially supported by the Center for Emergent Materials: an NSF MRSEC under award number DMR-1420451. The theoretical works are supported by the Graphene Flagship Core 2 grant number 785219 and Agence Nationale de la Recherche under references ANR-13-IS10-0003-01. This work was performed using HPC resources from GENCI-TGCC,CINES,IDRIS (Grant 2019-A0050901202 and *The Grand Challenge Jean Zay*).

Fig. 1. Image, electronic configuration and transport data of suspended r-TLG devices. Data taken at $T$=0.3K (a). Two-terminal differential conductance $G$ versus bias voltage at the CNP, showing a~40 meV gap. Inset: False-color SEM image of a suspended graphene device. (b). $G(n)$ at $V_b$=0. (c). Atomic and electronic configurations of r-TLG. Blue and red spheres indicate A-sublattice from the bottom layer and B-sublattice from the top layer, respectively. Grey spheres indicate stacked atoms. Arrows indicate spin directions. (d). $G(B_\parallel)$ at $V_{bg}$=13 V.

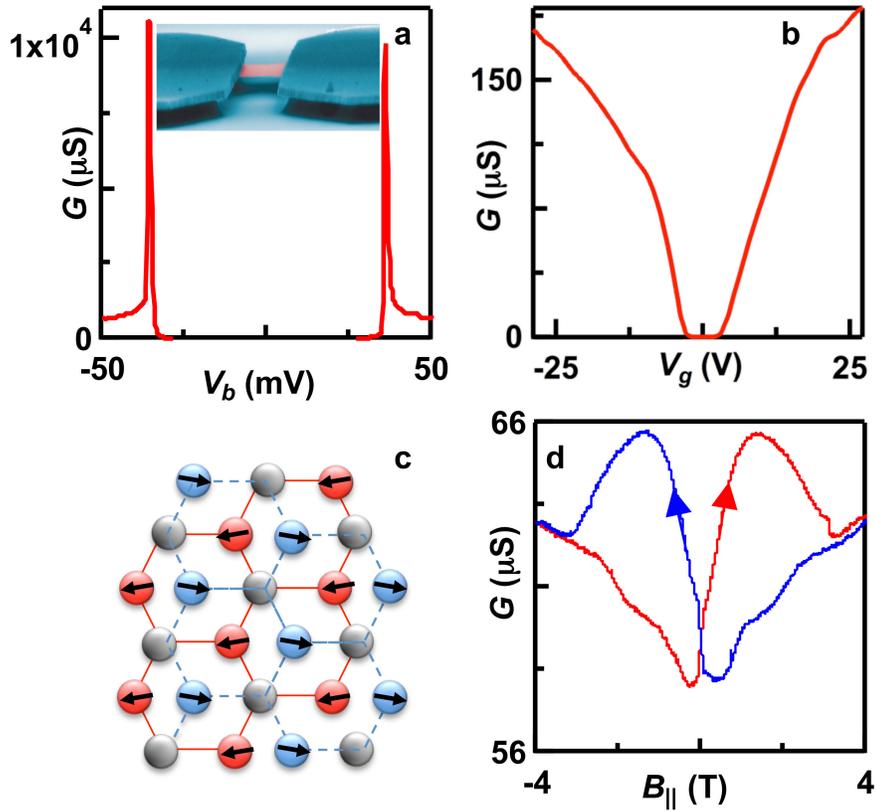

Fig. 2. Magnetoconductance and hysteresis in parallel and perpendicular magnetic fields of a r-TLG device that has gate capacitance of ~74 µF/m$^2$. (a). Magnetoconductance as a function of $B_\parallel$ at $T$=0.26K and charge densities $V_g$=-4.8, 12.2 and -11.3V, respectively. The curves are offset for clarity. (b). MC($B_\parallel$) at $V_g$=-3.8V and temperatures $T$=0.28, 0.4, 0.5, 0.6, 0.7K, 0.9, 1.9 and 4.2K (from blue to red). (c). MC($B_\perp$) at $T$=0.26K and $V_g$=-3.8V (purple), 10 (blue) and -10V, respectively. (d). MC($B_\perp$) at $V_g$=-4.6V and $T$=0.26K, 1.7K and 4.2K, respectively. In (c) and (d) the lower traces are multiplied by 3 and offset for better visibility.

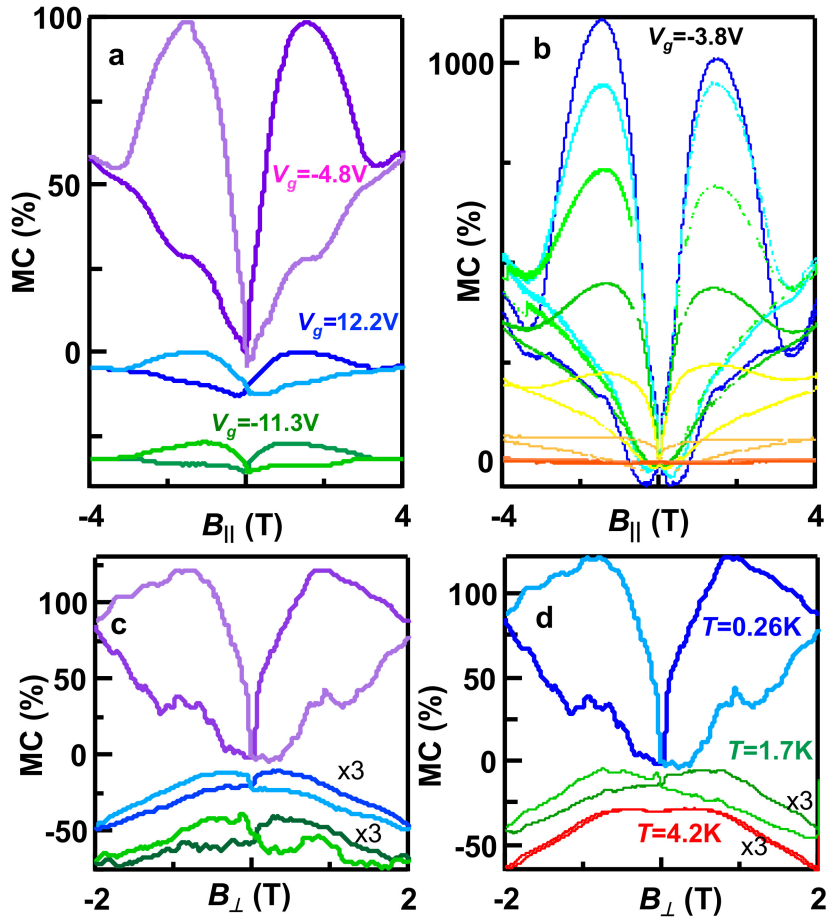

Fig. 3. Temperature dependence of MC and hysteresis. (a-b). MC at $V_g$=-3.8V as a function of temperature in linear and Arrhenius scales, respectively. (c). $\Delta G(T)$ at $V_g$=-3.8V and $B_{||}$=1.5T. (d). $B_{max}$, the $B_{||}$ value at which maximum $\Delta G$ is observed, as a funtion of $T$.

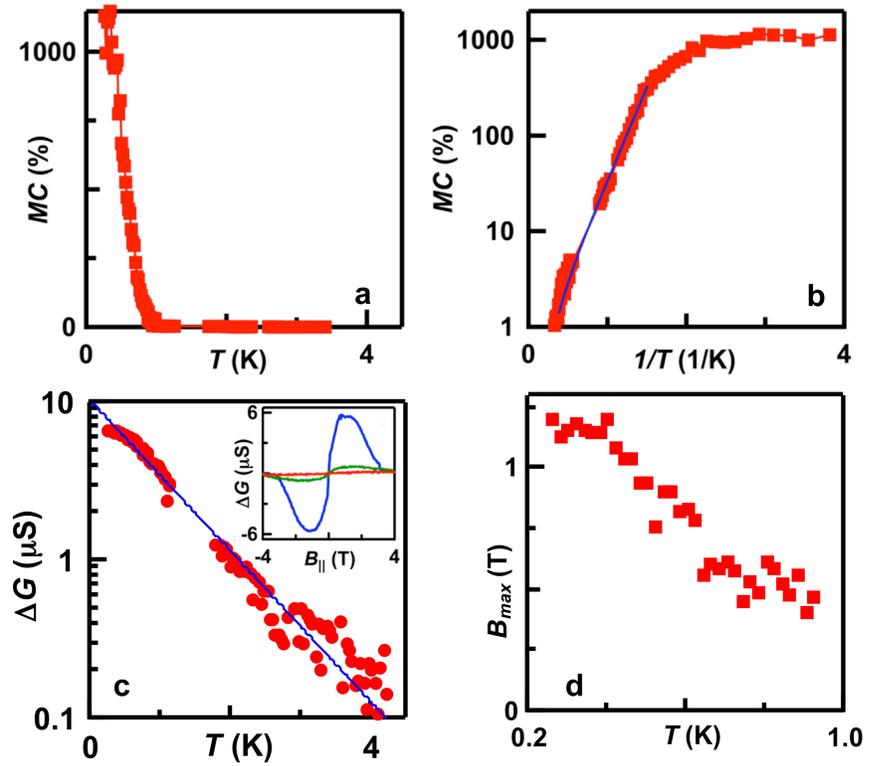

Fig. 4. Electronic structures of r-TLG and phase diagram. (a) Undoped r-TLG showing an insulating antiferromagnetic state. (b). a half-metallic solution at $n=-4\times10^{11}$ cm$^{-2}$, (c). a paramagnetic solution at $n=-12\times10^{11}$ cm$^{-2}$. (d). Calculated magnetization $m = \sum_{i=1}^{6} m_i$ (blue) and antiferromagnetic order parameter, $m = \sum_{i=1}^{6} |m_i|$, (red) as a function of n-type field effect doping. Here $|m_i|$ is the modulus of the magnetization on the $i^{th}$ atom in units of Bohr's magneton $\mu_B$. The magnetic half-metallic and paramagnetic phases are denoted in green and orange, respectively. In the calculation we assumed a collinear spin state. (e). Experimental data of MC vs $n$ at $B_\parallel$=1.65 T.

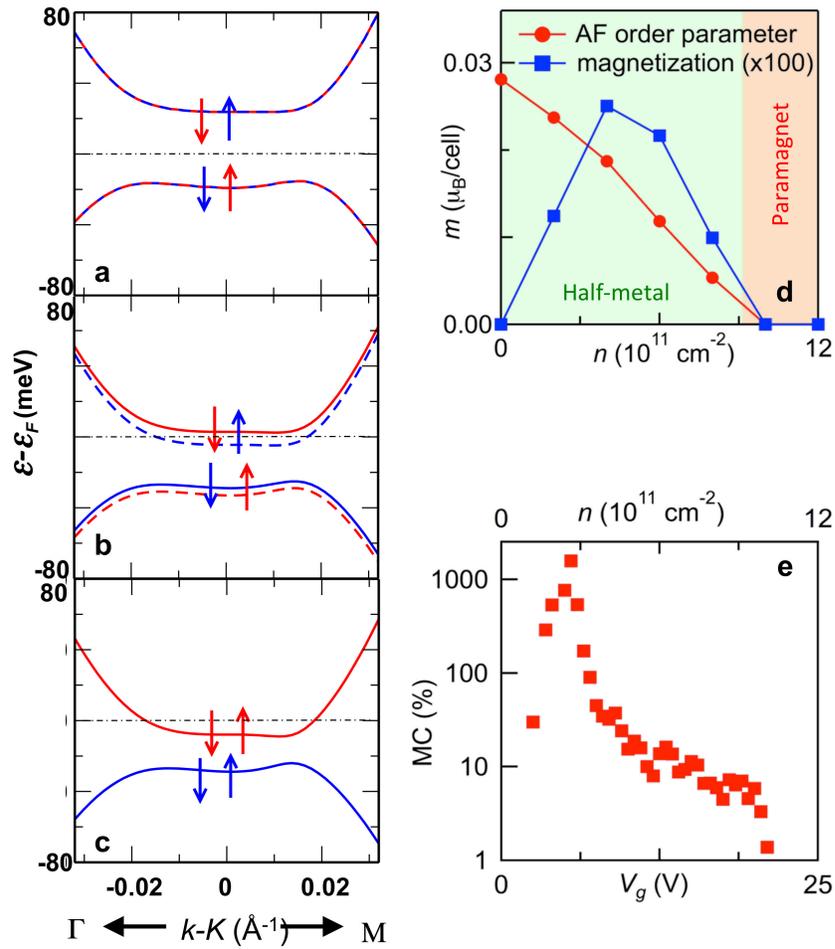